\documentclass{article}

\usepackage{xr}
\makeatletter

\newcommand*{\addFileDependency}[1]{
\typeout{(#1)}
\@addtofilelist{#1}
\IfFileExists{#1}{}{\typeout{No file #1.}}
}\makeatother

\newcommand*{\myexternaldocument}[1]{%
\externaldocument{#1}%
\addFileDependency{#1.tex}%
\addFileDependency{#1.aux}%
}

\myexternaldocument{article_SI}

\usepackage{arxiv}

\usepackage[utf8]{inputenc} 
\usepackage[T1]{fontenc}    
\usepackage{xr-hyper}
\usepackage{hyperref}       

\hypersetup{
    colorlinks=true,
    linkcolor=black,    
    urlcolor=black,     
    citecolor=black,    
    filecolor=black     
}

\usepackage{url}            
\usepackage{booktabs}       
\usepackage{amsmath,amsfonts,amssymb}       
\usepackage{nicefrac}       
\usepackage{microtype}      
\usepackage{graphicx}
\usepackage[square,numbers]{natbib}
\usepackage{doi}
\usepackage{siunitx}

\usepackage[acronym,shortcuts,automake=true]{glossaries}
\newacronym{FP}{FP}{Fabry-Pérot}
\newacronym{DBR}{DBR}{distributed Bragg reflector}
\newacronym{CFB}{CFB}{coherent fiber bundle}
\newacronym{SSE}{SSE}{spatio-spectral encoding}
\newacronym{FSR}{FSR}{free spectral range}
\newacronym{SEM}{SEM}{scanning electron microscope}
\newacronym{RCWA}{RCWA}{rigorous coupled-wave analysis}

\makeglossaries

\title{Spectrally-encoded non-scanning imaging through a fiber}

\author{Ningzhi Xie\\
	Department of Electrical and Computer Engineering\\
	University of Washington, Seattle, WA 98195 \\
	\texttt{nzxie@uw.edu} \\
	\And
	Quentin A. A. Tanguy \\
        Department of Electrical and Computer Engineering \\
        University of Washington, Seattle, WA 98195 \\
	\texttt{dr.quentin.tanguy@gmail.com} \\
	  \AND
	Johannes E. Fr{\"o}ch \\
	Department of Electrical and Computer Engineering\\
        Department of Physics \\
	University of Washington, Seattle, WA 98195 \\
	\texttt{jfroech@uw.edu } \\
        \And
	Karl F. B\"ohringer \\
	Department of Electrical and Computer Engineering\\
        Department of Bioengineering \\
	University of Washington, Seattle, WA 98195 \\
	\texttt{karlb@uw.edu} \\ 
 	  \And
	Arka Majumdar \\
	Department of Electrical and Computer Engineering\\
        Department of Physics \\
	University of Washington, Seattle, WA 98195 \\
	\texttt{arka@uw.edu} \\
}

\date{}

\hypersetup{
pdftitle={Spectrally-encoded non-scanning imaging through a fiber},
pdfsubject={optics},
pdfauthor={Ningzhi Xie},
pdfkeywords={Meta-optics, Imaging through fiber, spectral-encoding}
}

\begin{document}
\maketitle

\begin{abstract}
With the advent of neuroimaging and microsurgery, there is a rising need for capturing images through an optical fiber. We present an approach of imaging through a single fiber without mechanical scanning by implementing spatial-spectral encoding. The spectral encoding is achieved through a microfabricated spectral filter array, where light from different spatial pixels is coded with a highly orthogonal spectrum. The image is then computationally recovered via pseudo inverse of the encoding process. We demonstrate imaging of a $4 \times 4$ binary object at the proximity of the spectral filter array using $560-625nm$ wavelength band. The recovered image maintains an error rate of $<11\%$ when measured using a spectrometer with a spectral resolution of 1.5nm. The image remains unchanged with fiber bending or moving. Thus our approach shows a more robust way to image through a single optical fiber, with potential applications in compact endoscopes and angioscopes.
\end{abstract}


\section{Introduction}
Over the past half century, the possibility of imaging through an optical fiber has aroused an incessant strive to create novel methods and apparatuses. With the advent of neuroimaging and microsurgery, the need for miniaturization and flexibility for imaging devices has made optical fiber a key candidate. The practical ability to capture an image through an optical fiber can drastically transform the field of biomedical endoscopy. 

 Single-mode fibers can only transport intensities and phases of a single pixel due to the distortion of spatial distribution when transmitted through the core of the optical fiber. Transport of intensity or phase of multiple pixels, i.e., an image has been reported for fiber-optic confocal microscopy~\cite{FOCON1992}, scanning fiber endoscopy~\cite{SFE2010,Xie2022}, coherent fiber bundle~\cite{CFB1958} and MEMS-scanned optical coherence tomography endoscopic probe~\cite{Struk2018,Duan2016}. In scanning based modalities, an image is formed by scanning the fiber~\cite{SFE2010,Seo2018} or the beam at the distal end~\cite{Struk2018,Capasso2018metaOCT}, which results in bulky devices, relatively complicated integration and low imaging speed~\cite{Kim2019,Tanguy2020}. When using a large collection of single mode fibers (a coherent fiber bundle)~\cite{Miyoshi2023endoscope}, the resolution of the image is limited by the number of fibers per unit area in the bundle. 

 To transmit the spatial information of an image through a single fiber without mechanical scanning, various strategies have been extensively investigated using multi-mode fibers from as early as 1960's~\cite{spitz1967}. Several groups have exploited the fact that many modes can simultaneously exist and be transported in the core of multi-mode fibers, and demonstrate the possibility of imaging through a single fiber via index profile engineering and modal phase compensation. 
~\cite{yariv1976,friesem1983parallel}. Computational recovery of the distorted image through a multi-mode fiber has also been explored ~\cite{Choi2012}.
The fundamental idea is to model the propagation of the image through the multi-mode fiber using a transmission matrix, and computationally reconstruct the image by inverting the transmission matrix. However these methods lack robustness, and any disturbance or bending of the fiber necessitates re-calibration of the transmission matrix~\cite{Popoff2010}. More recently a promising approach \ac{SSE} has been reported ~\cite{BarankovSSE2014,KolenderskaSSE2015}. Here broadband light incident at different angles or positions is mapped to different spectra, which can be transmitted through a multi-mode fiber without any distortion. The spectra of light can be measured by a spectrometer at the back-end, followed by computational decoding to recover the distal spatial distribution of an object. Such encoding methods have been demonstrated by using two orthogonally tilted Fabry-P\'erot cavities~\cite{BarankovSSE2014} and a random scattering medium~\cite{KolenderskaSSE2015}.


However, in earlier works, the spectra encoded to the light incident from distinct spatial pixels, known as the spectral codes, were broad-band and non-designable. These codes contained high-frequency components that could only be captured through high-resolution ($\sim 0.5nm$ \cite{BarankovSSE2014}) spectral measurements. Furthermore, the codes were randomly generated, and hence one needs to fully rely on post-fabrication calibration to obtain them. This randomness also made the codes different for different devices, requiring calibration of every device, hindering ubiquitous practical application of \ac{SSE} in imaging devices. The randomized spectral codes also lacked orthogonality, leading to a low condition number for the spatial-spectral transfer matrix. As a result, the decoding process, which involved pseudo-inverting the transfer matrix, was sensitive to white noise and poorly characterized spectrum in the incoming light. Since the spectral codes were not designed, improving their orthogonality was challenging. Moreover, in these earlier works the \ac{SSE} optical devices are not microfabricated and compact, making integration with miniaturized imaging devices such as endoscopes difficult. 

\begin{figure}[ht!]
\centering
\includegraphics[width=0.95\textwidth]{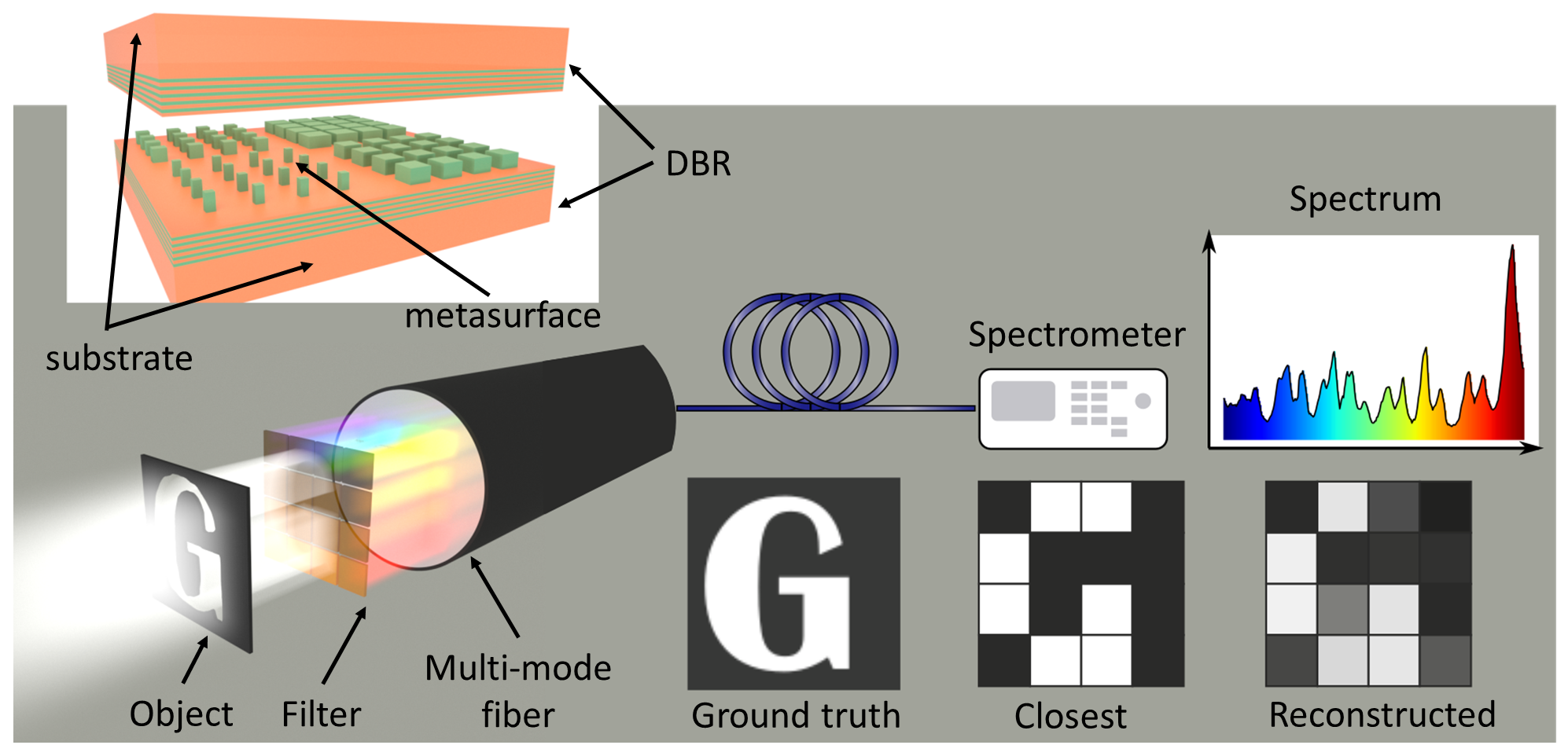}
\caption{Schematic of a SSE device based on a spectral filter array. \textbf{Inset:} Schematic of a metasurface phase-shifting layer wrapped in a \ac{FP} cavity.
} 
\label{fgr:Fig1_SENSE_sketch}
\end{figure}

SSE with designable and highly orthogonal spectral codes can be realized by creating an array of narrow band spectral filters with distinct passbands, each of which corresponds to the spectral code of one pixel of the SSE device, as shown in Fig.~\ref{fgr:Fig1_SENSE_sketch}.
Such a spectral filter array can be achieved by  embedding a transmissive dielectric metasurface as a phase shifting element in a \ac{FP} cavity, which was first demonstrated in a miniaturized infrared spectrometer~\cite{horie2016filter}. As shown in the inset of Fig.~\ref{fgr:Fig1_SENSE_sketch}, the \ac{FP} cavity is composed of two \acp{DBR} facing each other. The metasurface phase shifter increases the effective optical path length of the \ac{FP} cavity, red-shifting the transmitted resonant peak of the cavity. The amount of redshift is proportional to the amount of phase shift of the metasurface, which can be engineered by the lateral size of the meta-atoms. This DBR-metasurface-DBR spectral filter array is suitable for miniaturization and batch manufacturing. In this work, we design and fabricate a \ac{SSE} device based on this configuration, and perform spatial-spectral encoding and decoding experiments on $4 \times 4$ binary patterns in the wavelength range of $560-625nm$ using the fabricated device. By decoding using a high resolution ($\sim 0.2nm$) spectrometer, we recover the binary patterns with an average error rate of $9.8\%$ . We further show the error rate is maintained at $11\%$ even if the spectral resolution of the measurement drops to $\sim 1.6nm$.

\section{Results and Discussion}
\label{sec:Results}

\subsection*{Experimental Setup}
Our \ac{SSE} device is a 4$\times$4 pixel spectral filter array with a dimension of $50\mu m \times 50\mu m$ for each pixel. Each filter has a distinct passband (which are the spectral codes). A schematic of the metasurface-\ac{FP} cavity is provided in the lower right inset of Fig.~\ref{fgr:Fig2_setup_calib}a. Note that unlike the metasurface-\ac{FP} cavity filter array reported in the other paper \cite{horie2016filter}, we use SU8 as a spacer but not a cladding material. Our meta-atom stands in free space, because we use SiN as the meta-atom material for visible wavelength operation, which has a low refractive index of $\sim 2.0$. Embedding SiN meta-atom in SU8 will reduce the index contrast and hence the metasurface cannot impose enough phase delay to the light.

A 4$\times$4 pixel binary object is placed very close to the filter (distance $\sim 0.1mm$). 
To emulate the object, we used of a patterned chrome photomask placed at the focal plane of a focused laser beam (see Fig.~\ref{fgr:Fig2_setup_calib}a). The width of the beam in the focal plane is large enough to cover the full pattern. Each pixel of the binary object has the same width as that of the \ac{SSE} device. 
The light from the object passes through the \ac{SSE} device, which encode each spatial pixel into a unique spectral code. Then the spectrally encoded light is coupled into the optical fiber connected to a spectrometer to measure the transmitted spectrum. Detailed information of the setup can be found in the Method section.
Finally, the spectrum is computationally decoded to recover the pattern of the object using a pseudo inverse of the matrix \textbf{M} containing the superposed spectral codes. To minimize the cross-coupling between the spectral codes, the columns of the matrix \textbf{M} should be as orthogonal as possible. 

\subsection*{Design of the \ac{SSE} device}
\begin{figure}[ht!]
\begin{center}
\begin{tabular}{c}
\includegraphics[width=16cm]{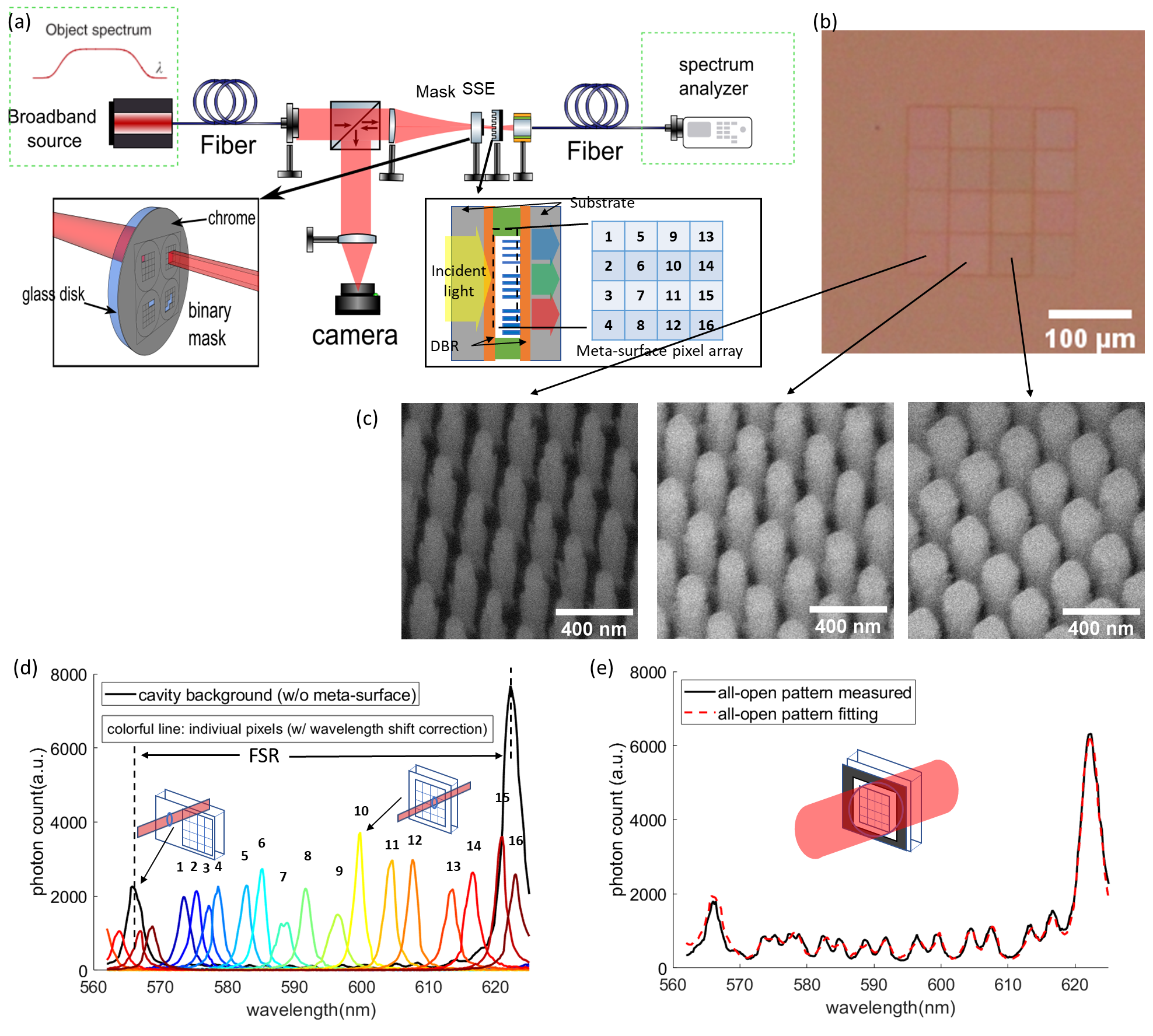}
\end{tabular}
\end{center}
\caption 
{\textbf{(a)} Schematic of the experimental setup for testing the spatial-to-spectral encoder. The lower left inserted figure shows the binary mask for creating patterns. The lower right inserted figure shows the cross-section of the SSE device. \textbf{(b)} The optical microscope image of the SSE device inspected from top. \textbf{(c)} The SEM images of the 3 metasurface embedded in the Fabry-Perot cavity at an oblique angle of $40^{\circ}$. The metasurface from left to right impose phase-shift of $\frac{1}{4} \pi$, $\frac{2}{4} \pi$, $\frac{3}{4} \pi$, respectively.  \textbf{(d)} The measured transmission spectra of 16 individual pixels of the SSE device (the spectral codes) in different colors from blue to red. These spectra are shifted to align to the corresponding resonance peaks of the transmission spectrum of the all-open pattern. The transmission spectrum of the cavity without metasurface (the background) is also plotted (in black) for comparison. \textbf{(e)} The measured and fitting spectra of the all-open pattern. The fitting spectrum is the weight summation of the spectra codes.} 
\label{fgr:Fig2_setup_calib}
\end{figure}

Ensuring orthogonality between the spectral codes is a key factor of the \ac{SSE} and can be achieved by distributing the \ac{FP} resonances as evenly as possible within the full \ac{FSR} of the \ac{FP}~\cite{BarankovSSE2014}. To form a resonant mode in a \ac{DBR}-metasurface-\ac{DBR} cavity, the round-trip phase $\psi$ must satisfy:
\begin{equation}
    \psi = 2(\phi_{i} + \frac{2\pi}{\lambda_{i}}L) = 2 \pi q
\end{equation}
where $\phi_{i}$ is the phase shift imparted by each metasurface, $\lambda_i$ is the corresponding resonance wavelength of each spectral filter, $L$ is the cavity length and $q$ is an integer, assuming the light impinging on the filter is a plane wave at normal incidence. We set $L = 2.5\mu m$, and use the resonance at $q = 9\gg1$. Thus, the transmission peak wavelength for each filter can be derived using a geometric sum approximation:
\begin{equation}
    \lambda_{i} = \frac{2L}{q-\phi_{i}/\pi} \approx \frac{2L}{q} + \frac{2L}{q^2}\frac{\phi_{i}}{\pi}
\end{equation}
which is around the design wavelength $\lambda_d = 560nm$. The \ac{FSR} is given by the difference between two adjacent orders and can be approximated as follows:
\begin{equation}
\Delta\lambda_\mathrm{FSR} = \frac{2L}{q-1} - \frac{2L}{q} \approx \frac{2L}{q^2}
\end{equation}
showing that in order to evenly cover the full \ac{FSR}, the phase shifts $\phi_{i}$ must evenly range from 0 to $\mathrm{\pi}$. We choose meta-atom material, height, and periodicity to cover the $0-\pi$ phase range at wavelength around $560nm$.The design and simulation of the metasurface can be found in the Method section. The simulated phase response of the meta-atom as a function of lateral size and wavelength are shown in the supporting information Fig.\ref{fgr:Fig_metaatom_response}. The fabrication details can be found in the Method section.  Fig.\ref{fgr:Fig2_setup_calib}b shows the optical image of the \ac{SSE} device inspected from top, while Fig.\ref{fgr:Fig2_setup_calib}c shows \ac{SEM} images of 3 different pixels of the metasurface, where the meta-atoms have different lateral size to impose different phase delays to the light.

\subsection*{Decoding}
The \ac{SSE} process can be described as $b$ = \textbf{M}$a$, where $a$ represents the $m \times 1$ vector of the input pixel values of a pattern, and b the $n \times 1 $ spectral output ~\cite{BarankovSSE2014}. \textbf{M} is the spectral encoding transfer matrix. Each column of \textbf{M} is the spectral code of one corresponding spatial pixel. Decoding consists in retrieving the input pattern $a$, which can be achieved by pseudo inverse of \textbf{M}: $a = $\textbf{M}$^{\dag} b$, where \textbf{M}$^{\dag}$ is the pseudo-inverse of matrix \textbf{M}. The details of the decoding algorithm can be found in the supporting information.

To retrieve the spatial input, \textbf{M} needs to be characterized prior to imaging. We measured the transmission spectra of all 16 single pixels of the \ac{SSE} device, which are defined as the spectral codes. We also measured the background, which is the transmission spectrum of the cavity when no metasurface is present. In our experiment we chose the wavelength range $560nm - 625nm$ for computational decoding, as this range is about one \ac{FSR} at the wavelength of $560nm$ for our \ac{FP} cavity that has a cavity length of $2.5\mu m$, and covers the resonance peaks of all 16 pixels (see Fig.~\ref{fgr:Fig_pixel_spec} in the SI). The measured spectral codes and background within this wavelength range are plotted in Fig.~\ref{fgr:Fig2_setup_calib}c. As can be seen in this figure, the peaks of the spectral codes are well separated with minimal overlap, indicating high orthogonality of the spectral codes. The peak of last two pixels overlap with the background as the metasurface phase shift of these two pixels is close to $\pi$. The effect of this overlap can be mitigated by including the background in the fitting in the decoding process. 

We note that the measured spectral codes undergo wavelength shift and amplitude modulation when compared to the spectrum of an all-open pattern, where all pixel intensities are defined as 1 (see Fig.~\ref{fgr:Fig_fullp_pixel} in the SI). Therefore, we calibrated the transfer matrix \textbf{M} by fitting the measured spectrum of the all-open pattern using the measured spectral codes and background. We shifted the measured spectral codes in wavelength to align to the resonance peaks of the spectrum of the all-open pattern, and add weight factors to modulate the relative amplitude of the spectral codes. The calibration details are in Section III of the SI. The fitted spectrum of the all-open pattern is plotted in Fig.~\ref{fgr:Fig2_setup_calib}d (red dash line), which matches well with the measured spectrum (black solid line). 

\begin{figure}[ht!]
\begin{center}
\begin{tabular}{c}
\includegraphics[width=16cm]{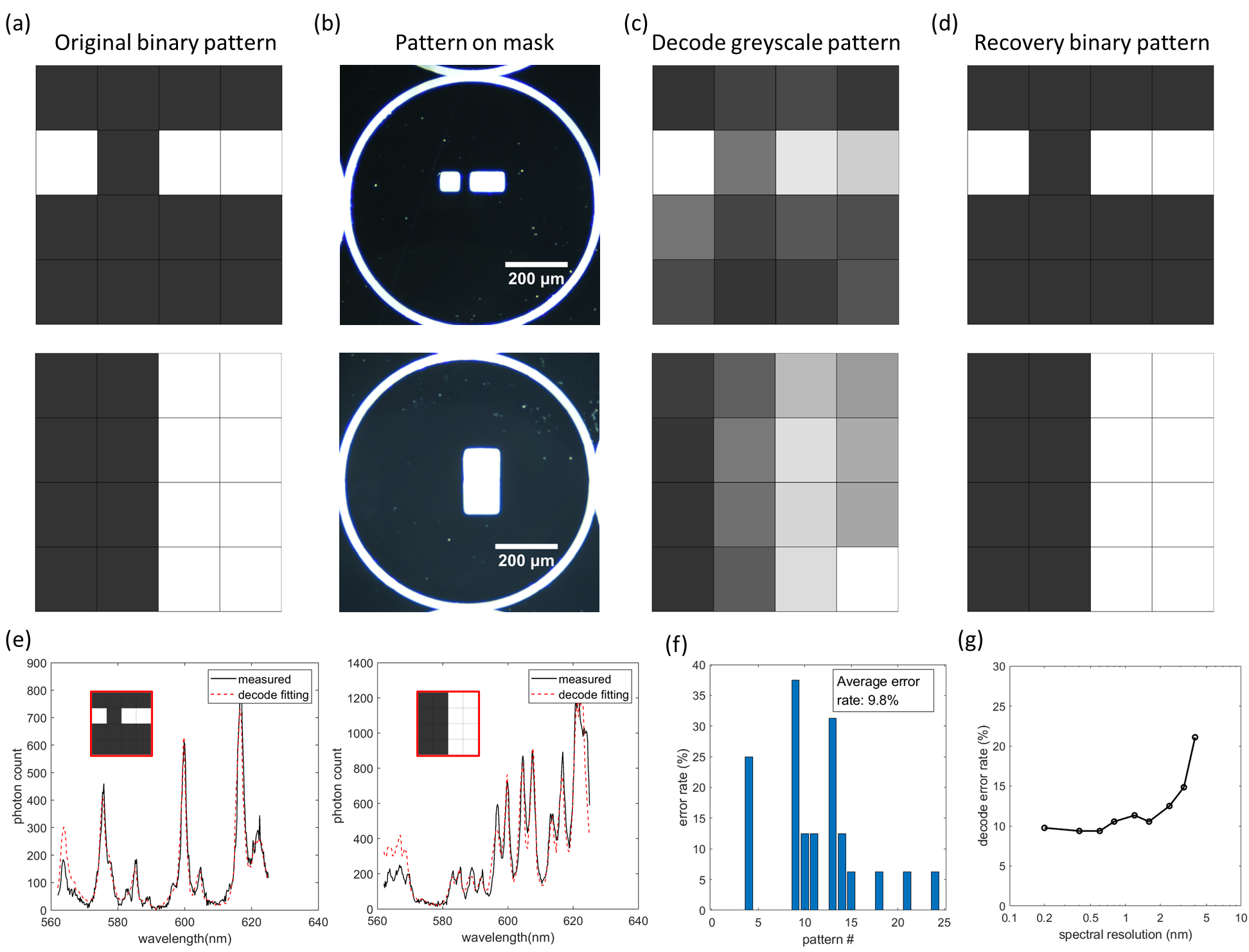}
\end{tabular}
\end{center}
\caption 
{\textbf{(a)} The $12^{\mathrm{th}}, 22^{\mathrm{th}}$ original binary patterns. \textbf{(b)} The microscope image of the corresponding binary patterns on the chrome mask. \textbf{(c)} The corresponding decoded grey-scale patterns. \textbf{(d)} The corresponding recovered binary pattern from the decoded grey-scale patterns. The threshold is set at the midpoint of the minimal and maximal greyscale values.  \textbf{(e)} The spectra of the corresponding patterns. The number of datapoints of the spectra is 316, for decoding 16 spatial pixels. \textbf{(f)} The error rates of the 24 recovered binary patterns compared to the original patterns, using the high resolution ($0.2nm$) measured spectra data, which contains $\approx 20$ spectrum datapoints per spatial pixels. \textbf{(g)} The pattern recovery average error rates (over 24 patterns) as a function of the resolution of the pattern spectra.}
\label{fgr:Fig_encode}
\end{figure}

Using the calibrated spectral encoding transfer matrix \textbf{M} and the decoding method described above, we recovered the binary patterns from the measured spectra of these patterns and compared them with the original patterns. Figs.~\ref{fgr:Fig_encode}a-e shows two examples of encoding and decoding of a binary pattern through our \ac{SSE} without error. The error rates of the encoding and decoding of all $24$ binary patterns are plotted in Fig.~\ref{fgr:Fig_encode}f. Here the decoding is peformed using the spectra measured by a high resolution ($0.2nm$) spectrometer, which contains $n=316$ data points in the wavelength range of $562-625nm$ ($20$ spectral data points per spatial pixel). This gives an average error rate of $9.8\%$. We investigated the dependency of the decoding error rate on the resolution of the pattern spectra. We synthesized low resolution spectral data by artificially compressing the number of data points of the measured pattern spectra and the spectral codes, and repeated the computational decoding on these low resolution pattern spectra.  As can be seen in Fig.~\ref{fgr:Fig_encode}g, the decoding error rate only marginally increases when the measured spectral resolution decreases, as long as the resolution is better than $1.6nm$. At $1.6nm$ resolution, the error rate is maintained at $\sim 11\%$. This shows that with our \ac{SSE}, the decoding can be efficiently performed using a low resolution spectrometer.  

\begin{figure}[ht!]
\begin{center}
\begin{tabular}{c}
\includegraphics[width=16cm]{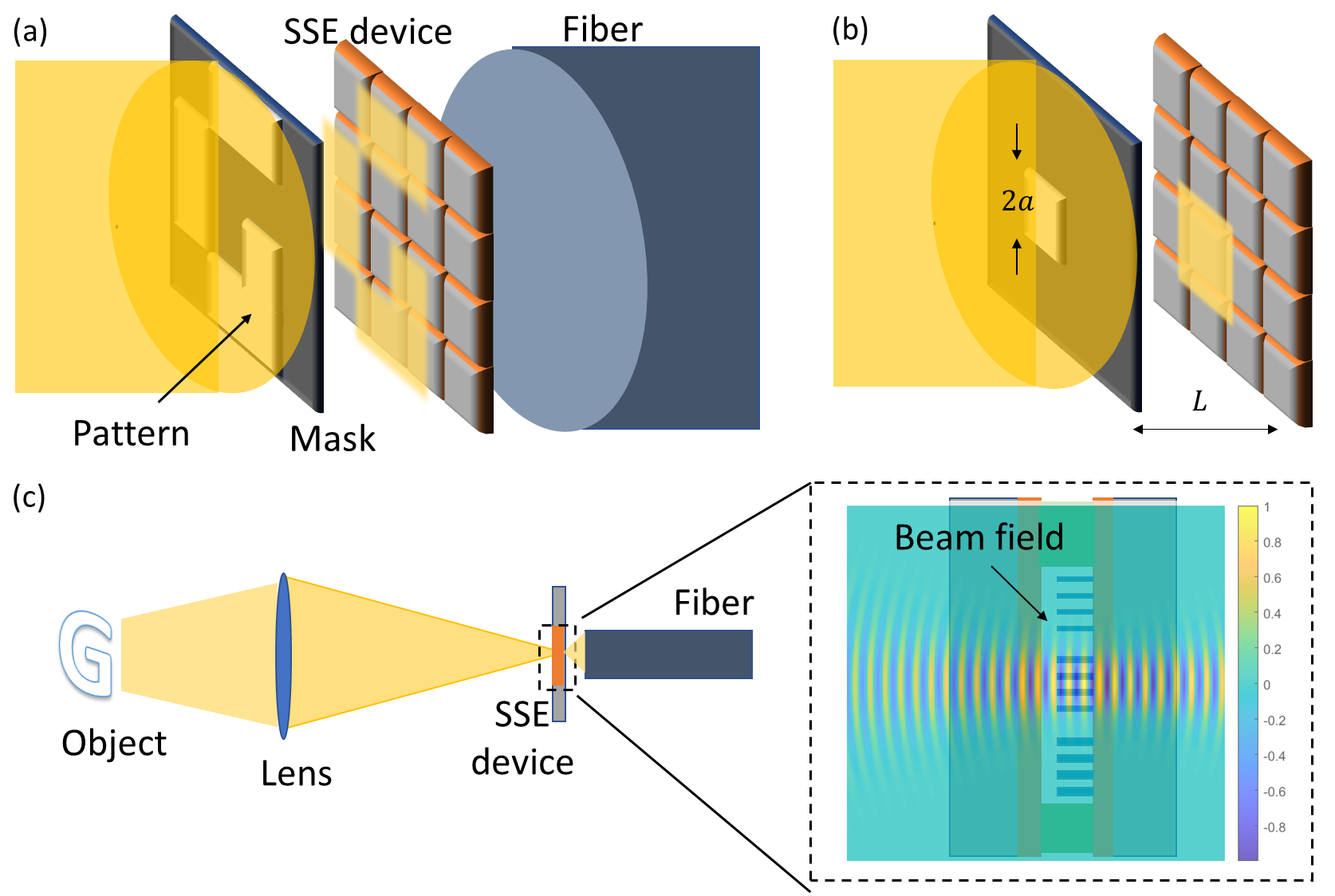}
\end{tabular}
\end{center}
\caption 
{\textbf{(a)} Pixel misalignment between the binary pattern on the mask and the SSE. \textbf{(b)} Pixel broadening in \ac{SSE} near field imaging. \textbf{(c)} Scheme for far-field imaging through a lens and the \ac{SSE} device.}
\label{fgr:Fig4_error}
\end{figure}

We attribute the $\sim 10\%$ average error rate of the spatial-spectral encoding and decoding experiment to the misalignment between the pixels of the pattern on the mask and the \ac{SSE} device, as demonstrated in Fig.~\ref{fgr:Fig4_error}a. Another challenge is the divergence of the light emitted from the pattern. Our \ac{SSE} device is based on the assumption that the incident light is a plane wave. This assumption is satisfied at near-field imaging condition, where the Fresnel number, $F=a^2/\lambda L \gg 1$. In our experiment setup, the characteristic length for a single pixel of the pattern is half the size of the pixel, $a=25\mu m$, the propagation distance $L\approx 600\mu m$ (slightly thicker than the substrate $500\mu m$, and the wavelength $\lambda\approx 560nm$. These parameters give a Fresnel number $F\approx 1.8$, which leads to the divergence of the light. As shown in Fig.~\ref{fgr:Fig4_error}b, the light from one single pixel of the pattern on the mask leaks to the surrounding pixels on the \ac{SSE} device due to the divergence of the light. The non-ideal plane wave incident of the light can cause deformations of the resonance transmission peaks of the cavity (the spectral codes of the \ac{SSE}), which may also result in errors in decoding.

The near field condition becomes more difficult to satisfy as the pixel size gets smaller, because the near field distance ($F>1$) is proportional to the square of the pixel size. In order to satisfy the near-field condition, for a pixel size of \SI{2}{\micro\meter} ($a$=\SI{1}{\micro\meter}), the working distance between the object and the filter must be smaller than \SI{1.8}{\micro\meter}, thus limiting the applications of \ac{SSE} in microscopy. On the other hand, as shown in Fig.~\ref{fgr:Fig4_error}c, the beam focused by a lens (Gaussian pilot beam) can be considered a plane wave near the focal point within a range defined by the Rayleigh length $z_R = \pi w_0^2/\lambda$. For a pixel size of \SI{2}{\micro\meter} ($w_0$=\SI{1}{\micro\meter}),  $z_R$=\SI{5.6}{\micro\meter}, which is larger than the thickness of the \ac{SSE} device. When the \ac{SSE} device is placed within this range, the plane wave condition is satisfied. Therefore, by using a focusing lens, we can potentially image in the far field via spectral encoding through the \ac{SSE} device even if the pixel size of the \ac{SSE} is reduced to $2\mu m$.

\section{Conclusion}
We designed and fabricated a \ac{SSE} device with controllable spectral code by creating an array of spectral filters, which is realized by wrapping a metasurface phase shifting layer in a \ac{FP} cavity. With this \ac{SSE} device, we perform a spatial-spectral encoding and decoding experiment on $4 \times 4$ pixel binary patterns (pixel size is $50\mu m\times 50 \mu m$). These patterns are created by a metal mask placed at the near field of the \ac{SSE} device, and are encoded into spectra by the device. The spectra of the patterns are decoded to recover the binary patterns by least-square fitting using the summation of the spectral codes of the \ac{SSE}. The decoded patterns have an average error rate of $9.7\%$ when the pattern spectra has a high resolution of $0.2nm$(the number of data points of spectra is $20\times$ the number of pixels of the patterns). This error rate is maintained at $\approx 11\%$ at much lower spectra resolution of $1.6nm$ (the spectral number of data point is $2.5\times$ the number of pixels), which means the spectra of the pattern can be measured by a miniaturized, low resolution spectrometer that is easier to be integrated on chip. 

Even though our encoding of the patterns are done at near field condition, far field imaging and encoding through our SSE device is possible by putting the SSE at the focal plane of a lens. For far field imaging, the size of the pixel of the SSE can possibly be reduced to $2 \mu m \times 2\mu m$. We also note that the number of pixels in our \ac{SSE} device, which is $16$ in this work, is limited by the \ac{FSR} of the \ac{FP} cavity overs the peak width of the spectral codes of the pixels (the resonance peak width), which is the Q factor of the cavity. Therefore, the number of pixels of the SSE is possible to be increased by making a higher Q cavity or using a photonic crystal spectral filter with high Q resonance \cite{Wang2019PCs}. 

\section{Method}
\label{sec:Methods}
\textbf{Experimental setup:} A broadband light source (Thorlab High-Power Stabilized Quartz Tungsten-Halogen Source: 360 - 2500 nm, SLS302) is coupled to an optical fiber, where the light emitted from the fiber is collimated by a tube lens (Thorlab AC254-030-A-ML, f = 30mm) and then focused by an objective to the chrome mask and/or the \ac{SSE} device. For measuring the transmission spectra of individual pixels of the \ac{SSE} device, the mask is not used, the fiber core diameter: \SI{10}{\micro\meter} (Thorlabs M64L01, 0.1NA), the objective: Nikon LU Plan Fluor 10x, $WD = 17.5mm, NA= 0.30$, the beam size on the SSE device: $\approx$ \SI{20}{\micro\meter}. For measuring the transmission spectra of the binary patterns on the mask, the fiber core diameter: \SI{200}{\micro\meter} (Thorlabs M122L02, 0.22NA), the objective: Nikon LU Plan Fluor 5x, $WD = 23.5mm, NA= 0.15$, the beam size on the mask: $\sim$\SI{800}{\micro\meter}. A 50/50 beam splitter (Not sure which one, there is no series number on the beam splitter) is placed in between the collimator and the objective, combined with a tube lens (Thorlabs AC254-100-A-ML, f=100mm) and a camera (Thorlabs CS165CU) for viewing the patterns on the mask. The binary patterns on the mask are created by photolithography. The light transmitted through the SSE device is coupled to another fiber (Thorlabs M74L05, core diameter = \SI{400}{\micro\meter}, 0.39NA), which is connected to a spectrometer (Teledyne Princeton Instruments, IsoPlane SCT 320) for measuring the spectrum of the light coming from the fiber.

\textbf{Design of the metasurface:} The metasurface consists of a 4$\times$4 pixel array, where each pixel is \SI{50}{}$\times$\SI{50}{\micro\meter} square. The metasurface is made of Si$_3$N$_4$ square meta-atoms on a fused silica substrate (height=\SI{500}{\nano\meter}, square lattice, pitch=\SI{300}{\nano\meter}). The size of the meta-atoms for each individual pixel is selected to satisfy the phase shift requirement: for the $\textit{i}\,^{\mathrm{th}}$ pixel, the (single pass) phase shift is $\frac{i}{16} \pi$ at the resonance wavelength of the \ac{FP} cavity of that pixel. We use \ac{RCWA}~\cite{S4} to theoretically calculate the phase response as a function of the meta-atom lateral size and wavelength. The results are shown in the supporting information Fig.\ref{fgr:Fig_metaatom_response}

 These 2 DBRs are bonded together with a spacing of \SI{2.5}{\micro\meter}, which defines the cavity length $L$, using the SU8 resist as the bonding material and the spacer.

\textbf{Fabrication of the SSE device:}  The two DBRs are grown on two \SI{500}{\micro\meter}-thick fused-silica substrates (\SI{10}{}$\times$\SI{10}{\milli\meter}$^2$ chip) with plasma-enhanced chemical vapor deposition. These DBRs are made of 7 alternating layers of Si$_3$N$_4$ ($n=1.97$ at \SI{560}{\nano\meter}, thickness = \SI{71}{\nano\meter}) and SiO$_2$ ($n=1.44$ at \SI{560}{\nano\meter}, thickness = \SI{97}{\nano\meter}), which has a reflecting band centered at \SI{560}{\nano\meter}. After that, \SI{500}{\nano\meter} Si$_3$N$_4$ is deposited on one of the DBRs, where the metasurface is fabricated by electron beam lithography (EBL).  A \SI{300}{\nano\meter}-thick e-beam resist (ZEP-520A) is spin-coated and patterned by EBL, then a $\approx$ \SI{65}{\nano\meter} thick Al$_2$O$_3$ hard mask is created by electron beam assisted evaporation followed by resist lift-off in N-methyl-2-pyrrolidone (NMP) at $90^\circ$C overnight. Subsequently, the Si$_3$N$_4$ layer is etched by a fluorine based reactive ion process. Meanwhile, a \SI{2.5}{\micro\meter}-thick SU8 resist is spin-coated on the other DBR and patterned by photolithography to create a spacer structure (A ring with inner diameter = \SI{5}{\milli\meter}, outer diameter = \SI{8}{\milli\meter}, height = \SI{2.5}{\micro\meter}). Finally, the two DBRs are bonded together with metasurface wrapped inside using a flip-chip bonder. The SU8 bonding is done by heating up the two substrates to $250^{\circ}$C and applying \SI{40}{\newton} force for 1 hour. The sketch of the process flow can be seen in Fig.\ref{fgr:Fig_processflow} in the supporting information.

\section*{Disclosures}
Arka Majumdar and Karl F. Böhringer are co-founder of Tunoptix, which is looking into commercializing metaoptics. 

\section*{Acknowledgements}
The work is supported by NSF GCR 2120774. Part of this work was conducted at the Washington Nanofabrication Facility / Molecular Analysis Facility, a National Nanotechnology Coordinated Infrastructure (NNCI) site at the University of Washington with partial support from the National Science Foundation (NSF) via awards NNCI-1542101 and NNCI-2025489.

\vspace{12pt}

\bibliographystyle{abbrvnat}
\bibliography{references}  

\begin{thebibliography}{21}
\providecommand{\natexlab}[1]{#1}
\providecommand{\url}[1]{\texttt{#1}}
\expandafter\ifx\csname urlstyle\endcsname\relax
  \providecommand{\doi}[1]{doi: #1}\else
  \providecommand{\doi}{doi: \begingroup \urlstyle{rm}\Url}\fi

\bibitem[Barankov and Mertz(2014)]{BarankovSSE2014}
R.~Barankov and J.~Mertz.
\newblock High-throughput imaging of self-luminous objects through a single
  optical fibre.
\newblock \emph{Nature Communications}, 5, 2014.
\newblock ISSN 2041-1723.
\newblock \doi{ARTN 5581 10.1038/ncomms6581}.
\newblock URL \url{<Go to ISI>://WOS:000345740900003}.

\bibitem[Choi et~al.(2012)Choi, Yoon, Kim, Yang, Fang-Yen, Dasari, Lee, and
  Choi]{Choi2012}
Y.~Choi, C.~Yoon, M.~Kim, T.~D. Yang, C.~Fang-Yen, R.~R. Dasari, K.~J. Lee, and
  W.~Choi.
\newblock Scanner-free and wide-field endoscopic imaging by using a single
  multimode optical fiber.
\newblock \emph{Physical Review Letters}, 109\penalty0 (20), 2012.
\newblock ISSN 0031-9007.
\newblock \doi{ARTN 203901 10.1103/PhysRevLett.109.203901}.
\newblock URL \url{<Go to ISI>://WOS:000310978500004}.
\newblock 035vi Times Cited:376 Cited References Count:30.

\bibitem[Dabbs and Glass(1992)]{FOCON1992}
T.~Dabbs and M.~Glass.
\newblock Fiber-optic confocal microscope: Focon.
\newblock \emph{Appl Opt}, 31\penalty0 (16):\penalty0 3030--5, 1992.
\newblock ISSN 1559-128X (Print) 1559-128X (Linking).
\newblock \doi{10.1364/AO.31.003030}.
\newblock URL \url{https://www.ncbi.nlm.nih.gov/pubmed/20725247}.

\bibitem[Duan et~al.(2016)Duan, Tanguy, Pozzi, and Xie]{Duan2016}
C.~Duan, Q.~Tanguy, A.~Pozzi, and H.~Xie.
\newblock Optical coherence tomography endoscopic probe based on a tilted mems
  mirror.
\newblock \emph{Biomedical Optics Express}, 7, 2016.
\newblock \doi{10.1364/boe.7.003345}.

\bibitem[Friesem et~al.(1983)Friesem, Levy, and
  Silberberg]{friesem1983parallel}
A.~Friesem, U.~Levy, and Y.~Silberberg.
\newblock Parallel transmission of images through single optical fibers.
\newblock \emph{Proceedings of the IEEE}, 71\penalty0 (2):\penalty0 208--221,
  1983.

\bibitem[Hirschowitz et~al.(1958)Hirschowitz, Curtiss, Peters, and
  Pollard]{CFB1958}
B.~I. Hirschowitz, L.~E. Curtiss, C.~W. Peters, and H.~M. Pollard.
\newblock Demonstration of a new gastroscope, the fiberscope.
\newblock \emph{Gastroenterology}, 35\penalty0 (1):\penalty0 50--53, 1958.
\newblock ISSN 0016-5085.
\newblock URL \url{<Go to ISI>://WOS:A1958WA70000006}.
\newblock Wa700 Times Cited:220 Cited References Count:1.

\bibitem[Horie et~al.(2016)Horie, Arbabi, Arbabi, Kamali, and
  Faraon]{horie2016filter}
Y.~Horie, A.~Arbabi, E.~Arbabi, S.~M. Kamali, and A.~Faraon.
\newblock {Wide bandwidth and high resolution planar filter array based on
  DBR-metasurface-DBR structures}.
\newblock \emph{Optics Express}, 24\penalty0 (11):\penalty0 11677--11682, 2016.
\newblock ISSN 1094-4087.
\newblock \doi{10.1364/oe.24.011677}.

\bibitem[Kim et~al.(2019)Kim, Hwang, Ahn, Seo, Kim, Lee, Yoon, Kong, Jeong,
  Jon, Kim, and Jeong]{Kim2019}
D.~Y. Kim, K.~Hwang, J.~Ahn, Y.~H. Seo, J.~B. Kim, S.~Lee, J.~H. Yoon, E.~Kong,
  Y.~Jeong, S.~Jon, P.~Kim, and K.~H. Jeong.
\newblock Lissajous scanning two-photon endomicroscope for in vivo tissue
  imaging.
\newblock \emph{Scientific Reports}, 9, 12 2019.
\newblock ISSN 20452322.
\newblock \doi{10.1038/s41598-019-38762-w}.

\bibitem[Kolenderska et~al.(2015)Kolenderska, Katz, Fink, and
  Gigan]{KolenderskaSSE2015}
S.~M. Kolenderska, O.~Katz, M.~Fink, and S.~Gigan.
\newblock Scanning-free imaging through a single fiber by random
  spatio-spectral encoding.
\newblock \emph{Optics Letters}, 40\penalty0 (4):\penalty0 534--537, 2015.
\newblock ISSN 0146-9592.
\newblock \doi{10.1364/Ol.40.000534}.
\newblock URL \url{<Go to ISI>://WOS:000349848400025}.
\newblock Cb7zu.

\bibitem[Lee et~al.(2010)Lee, Engelbrecht, Soper, Helmchen, and
  Seibel]{SFE2010}
C.~M. Lee, C.~J. Engelbrecht, T.~D. Soper, F.~Helmchen, and E.~J. Seibel.
\newblock Scanning fiber endoscopy with highly flexible, 1 mm catheterscopes
  for wide-field, full-color imaging.
\newblock \emph{J Biophotonics}, 3\penalty0 (5-6):\penalty0 385--407, 2010.

\bibitem[Liu and Fan(2012)]{S4}
V.~Liu and S.~Fan.
\newblock S4: A free electromagnetic solver for layered periodic structures.
\newblock \emph{Computer Physics Communications}, 183:\penalty0 2233--2244, 10
  2012.
\newblock ISSN 00104655.
\newblock \doi{10.1016/j.cpc.2012.04.026}.

\bibitem[Miyoshi et~al.(2023)Miyoshi, Nishimura, Shimojo, Okayama, and
  Awazu]{Miyoshi2023endoscope}
Y.~Miyoshi, T.~Nishimura, Y.~Shimojo, K.~Okayama, and K.~Awazu.
\newblock Endoscopic image‑guided laser treatment system based on fiber
  bundle laser steering.
\newblock \emph{Scientific Reports}, 13:\penalty0 2921, 2023.
\newblock \doi{10.1038/s41598-023-29392-4}.
\newblock URL \url{https://doi.org/10.1038/s41598-023-29392-4}.

\bibitem[Pahlevaninezhad et~al.(2018)Pahlevaninezhad, Khorasaninejad, and
  Huang]{Capasso2018metaOCT}
H.~Pahlevaninezhad, M.~Khorasaninejad, and Y.-W. Huang.
\newblock Nano-optic endoscope for high-resolution optical coherence tomography
  in vivo.
\newblock \emph{nature Photonics}, 12:\penalty0 540--547, 9 2018.
\newblock \doi{10.1038/s41566-018-0224-2}.
\newblock URL \url{https://doi.org/10.1038/s41566-018-0224-2}.

\bibitem[Popoff et~al.(2010)Popoff, Lerosey, Carminati, Fink, Boccara, and
  Gigan]{Popoff2010}
S.~M. Popoff, G.~Lerosey, R.~Carminati, M.~Fink, A.~C. Boccara, and S.~Gigan.
\newblock Measuring the transmission matrix in optics: An approach to the study
  and control of light propagation in disordered media.
\newblock \emph{Physical Review Letters}, 104:\penalty0 100601, 3 2010.
\newblock ISSN 00319007.
\newblock \doi{10.1103/PhysRevLett.104.100601}.

\bibitem[Seo et~al.(2018)Seo, Hwang, and Jeong]{Seo2018}
Y.-H. Seo, K.~Hwang, and K.-H. Jeong.
\newblock 165 mm diameter forward-viewing confocal endomicroscopic catheter
  using a flip-chip bonded electrothermal mems fiber scanner.
\newblock \emph{Optics Express}, 26\penalty0 (4):\penalty0 4780, 2018.
\newblock \doi{10.1364/oe.26.004780}.

\bibitem[Spitz and Werts(1967)]{spitz1967}
E.~Spitz and A.~Werts.
\newblock Transmission des images {\`a} travers une fibre optique.
\newblock \emph{C. R. Acad. Sci. B}, 264\penalty0 (14):\penalty0 1015, 1967.

\bibitem[Struk et~al.(2018)Struk, Bargiel, Tanguy, Ramirez, Passilly, Lutz,
  Gaiffe, Xie, and Gorecki]{Struk2018}
P.~Struk, S.~Bargiel, Q.~A.~A. Tanguy, F.~E.~G. Ramirez, N.~Passilly, P.~Lutz,
  O.~Gaiffe, H.~Xie, and C.~Gorecki.
\newblock Swept-source optical coherence tomography microsystem with an
  integrated mirau interferometer and electrothermal micro-scanner.
\newblock \emph{Optics Letters}, 43:\penalty0 4847, 10 2018.
\newblock ISSN 0146-9592.
\newblock \doi{10.1364/ol.43.004847}.

\bibitem[Tanguy et~al.(2020)Tanguy, Gaiffe, Passilly, Cote, Cabodevila,
  Bargiel, Lutz, Xie, and Gorecki]{Tanguy2020}
Q.~A.~A. Tanguy, O.~Gaiffe, N.~Passilly, J.-M. Cote, G.~Cabodevila, S.~Bargiel,
  P.~Lutz, H.~Xie, and C.~Gorecki.
\newblock Real-time lissajous imaging with a low-voltage 2-axis mems scanner
  based on electrothermal actuation.
\newblock \emph{Optics Express}, 28:\penalty0 8512, 3 2020.
\newblock \doi{https://doi.org/10.1364/OE.380690}.

\bibitem[Wang et~al.(2019)Wang, Yi, Chen, Zhou, Luk, James, Nogan, Ross, Joe,
  Shahsafi, Wang, Kats, and Yu]{Wang2019PCs}
Z.~Wang, S.~Yi, A.~Chen, M.~Zhou, T.~S. Luk, A.~James, J.~Nogan, W.~Ross,
  G.~Joe, A.~Shahsafi, K.~X. Wang, M.~A. Kats, and Z.~F. Yu.
\newblock Single-shot on-chip spectral sensors based on photonic crystal slabs.
\newblock \emph{Nature Communications}, 10, 2019.
\newblock ISSN 2041-1723.
\newblock \doi{ARTN 1020 10.1038/s41467-019-08994-5}.
\newblock URL \url{<Go to ISI>://WOS:000460125400002}.
\newblock Hn4aa Times Cited:147 Cited References Count:49.

\bibitem[Xie et~al.(2022)Xie, Carson, Fröch, Majumdar, Seibel, and
  Böhringer]{Xie2022}
N.~Xie, M.~D. Carson, J.~E. Fröch, A.~Majumdar, E.~Seibel, and K.~F.
  Böhringer.
\newblock Large fov short-wave infrared meta-lens for scanning fiber endoscopy.
\newblock \emph{Journal of Biomedical Optics}, 28:\penalty0 094802, 12 2022.
\newblock ISSN 15602281.
\newblock \doi{10.1117/1.JBO.28.9.094802}.
\newblock URL \url{http://arxiv.org/abs/2212.11272}.

\bibitem[Yariv(1976)]{yariv1976}
A.~Yariv.
\newblock Three-dimensional pictorial transmission in optical fibers.
\newblock \emph{Applied Physics Letters}, 28\penalty0 (2):\penalty0 88--89,
  1976.

\end{thebibliography}

\newpage

\section*{Supporting information}


\renewcommand{\thefigure}{S\arabic{figure}}
\setcounter{figure}{0}

\subsection*{I. Metasurface design}

\begin{figure}[ht!]
    \begin{center}
        \begin{tabular}{c}
        \includegraphics[width=16cm]{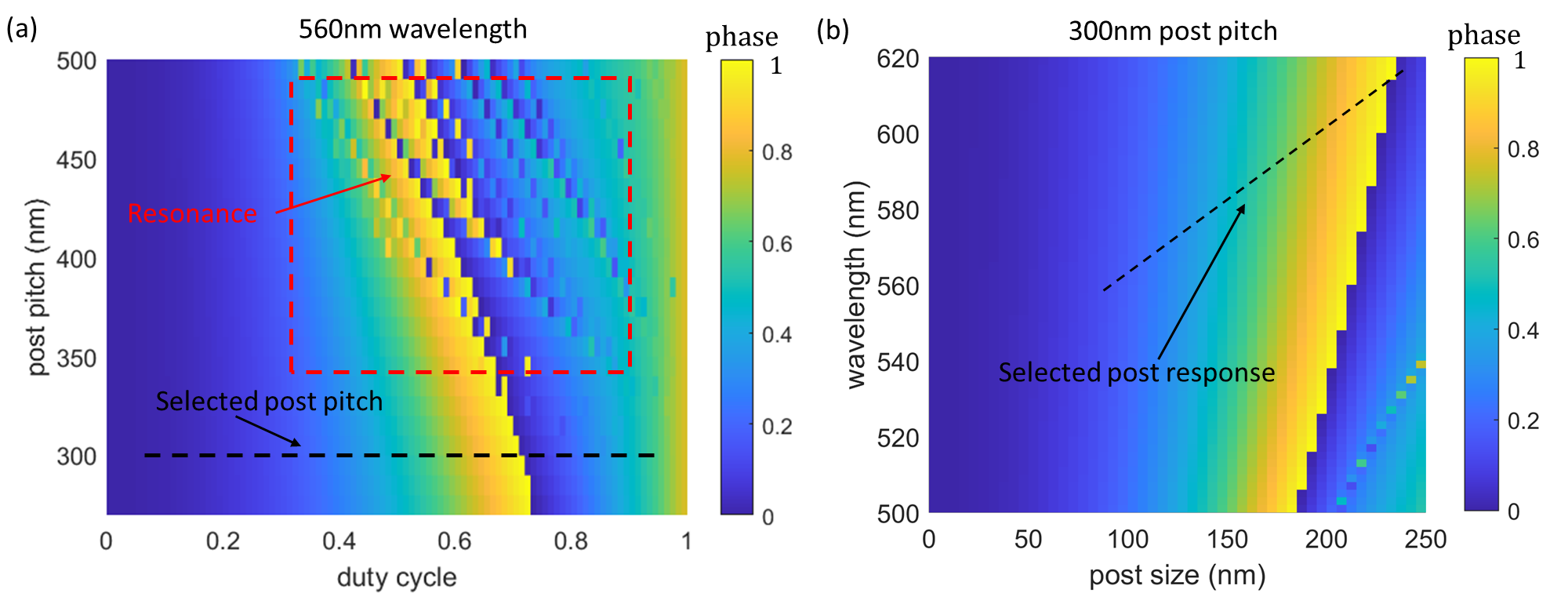}
        \end{tabular}
    \end{center}
    \caption{\textbf{(a)} Square nano-post meta-atom phase response as function of post pitch and duty cycle (post lateral size divided by post pitch) simulated by RCWA. The phase is normalized and wrapped by $\pi$. The nano-posts are made of SiN, on square lattice, and have a height of $500nm$. The simulation is done at the wavelength of 560nm. \textbf{(b)} The meta-atom phase response as a function of wavelength and post size.} 
    \label{fgr:Fig_metaatom_response}
\end{figure}

To have a meta-atom library that cover the $\pi$ phase range at around $560nm$ wavelength, we first perform a parameter sweep on the meta-atom pitch and lateral size. The phase response of the meta-atom is plotted in Fig.\ref{fgr:Fig_metaatom_response}a. We select a meta-atom pitch of $300nm$. At this pitch the phase increase smoothly and continuously with the meta-atom size, which indicates no resonance mode exist in the meta-atom array. We choose meta-atom lateral size ranging from $90nm$ to $240nm$, which has a weak wavelength dependent phase response, and covers the $\pi$ phase range from $560nm$ to $620nm$. As can be seen in Fig.\ref{fgr:Fig_metaatom_response}b. The phase response of the meta-atoms at the resonance wavelength of the FP cavity are indicated by a black dash line.  

\newpage

\subsection*{II. Measured spectra of SSE pixels and encoded patterns}

\begin{figure}[ht!]
    \begin{center}
        \begin{tabular}{c}
        \includegraphics[width=16cm]{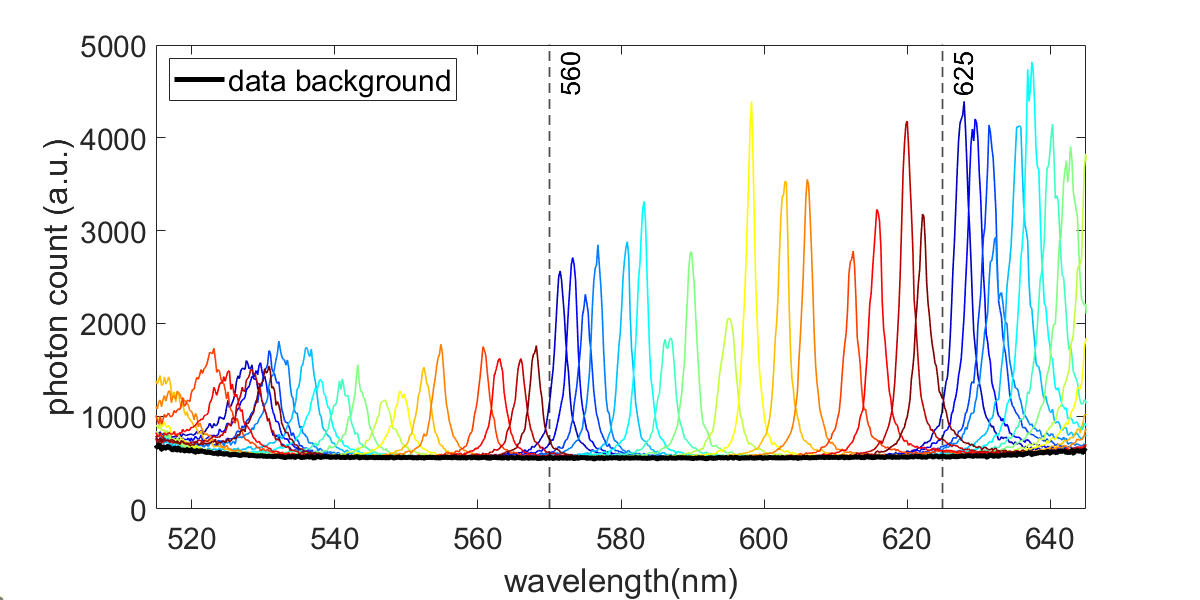}
        \end{tabular}
    \end{center}
    \caption{Raw data of the measured transmission spectra of the broad band income light passing through 16 different single pixels of the spectral encoder. The color from blue to red represents pixel 1 to 16} 
    \label{fgr:Fig_pixel_spec}
\end{figure} 

\begin{figure}[ht!]
    \begin{center}
        \begin{tabular}{c}
        \includegraphics[width=16cm]{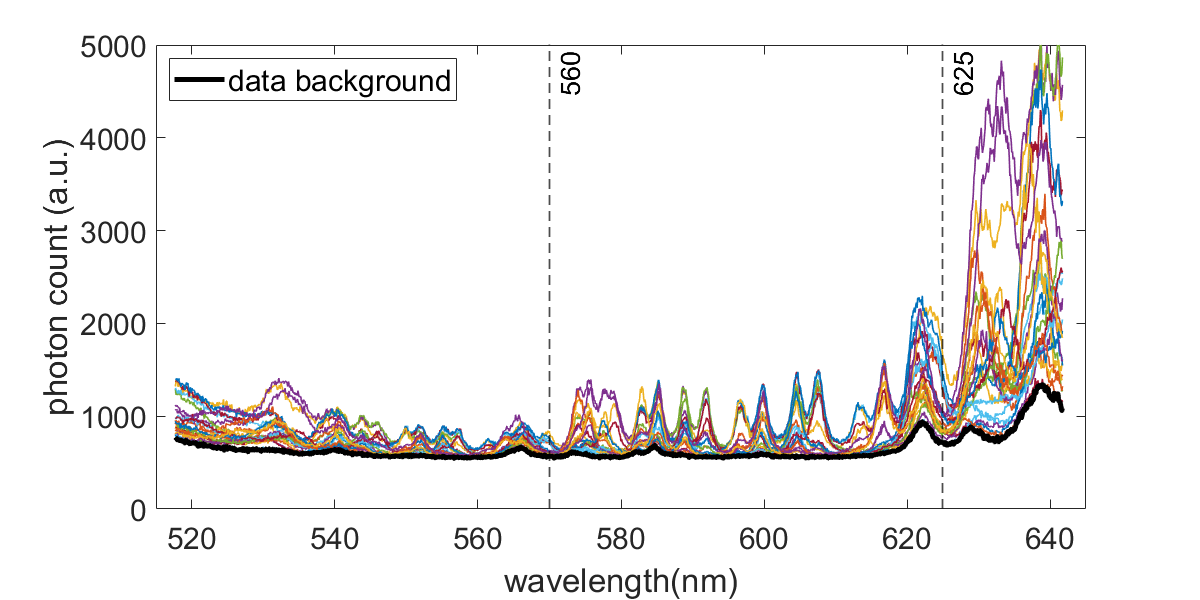}
        \end{tabular}
    \end{center}
    \caption{Raw data of the measured transmission spectra of the broad band income light passing through 24 different binary patterns and the spectral encoder} 
    \label{fgr:Fig_pattern_spec}
\end{figure} 

\newpage

\subsection*{III. Decoding algorithm}

The spectrum of a given pattern, $S_{pat}(\lambda)$ (The matrix $B$) is a weighted summation of the spectra of the $16$ individual pixels, $S_i(\lambda)$, and the background, $S_{bg}(\lambda)$(the columns of \textbf{M}):

\begin{equation}\label{eq:Encode}
\underbrace{\begin{bmatrix}
S_{pat}(\lambda)
\end{bmatrix}}_{b}
 = 
\underbrace{\begin{bmatrix}
S_{bg}(\lambda) & S_1(\lambda) & \ldots & S_{16}(\lambda)
\end{bmatrix}}_{\textbf{M}}
\cdot
\underbrace{\begin{bmatrix}
a_0 \\ a_1 \\ \vdots \\ a_{16}
\end{bmatrix}}_{a}
\end{equation}

where $a_{i}$ is the intensity value of the $\textit{i}^{\mathrm{th}}$ pixel (The element in the vector $a$). 

The decoding consists in fitting the $\{a_{i}\}$ coefficients to minimize the square error of $S_{pat}(\lambda)$ compared to the measured spectrum $S_{pat-m}(\lambda)$ as defined in Eq.~\ref{eq:Decode} where $[\lambda_1,\lambda_2]$ is the wavelength range of interest. This fitting can be done by finding the pseudo-inverse of matrix \textbf{M}.

\begin{align}
\label{eq:Decode}
 \min_{a_{i}} \parallel S_{pat}(\lambda) - S_{pat-m}(\lambda) \parallel^2 &= \min_{a_{i}} \int_{\lambda_1}^{\lambda_2} |a_{0}| S_{bg}(\lambda) + \sum_{i=1}^{M} a_{i} S_i(\lambda) - S_{pat-m}(\lambda)|^2 \, d\lambda \\
 \Rightarrow \min_{a} \parallel Ma-b_{meas} \parallel^2 
 \Rightarrow a &= M^{+} b_{meas} 
 \text{      where  } M^{+} = (M^T M)^{-1} M^T
\end{align}

The range $[\lambda_1,\lambda_2]$ is chosen from the measured spectra of individual pixels of the SSE (see Fig.~\ref{fgr:Fig_pixel_spec}), which is $[560nm,625nm]$. 

\newpage

\subsection*{IV. Spectral code calibration}

We calibrate the spectral code matrix \textbf{M} by fitting the measured spectrum of the all-open pattern $S_{apat-m}(\lambda)$ using the measured spectral codes and background, $S_{i-m}(\lambda)$, $S_{bg-m}(\lambda)$:

\begin{align}
b_{apat} &= \begin{bmatrix} S_{apat-m}(\lambda)\end{bmatrix} \\
M &= \begin{bmatrix}
 c_0 S_{bg-m}(\lambda+\Delta \lambda_0) & c_1S_{1-m}(\lambda + \Delta \lambda_1) & \ldots & c_{16} S_{16-m}(\lambda + \Delta \lambda_{16})
\end{bmatrix} \\
a_{apat} &= \begin{bmatrix}
1 & 1 & \ldots & 1 \end{bmatrix}^T \\
\min_{c_i,\Delta \lambda_i} &\parallel  M a_{apat} - b_{apat} \parallel^2
\end{align}


where $c_i$ are weight factors and $\Delta \lambda_i$ are the peak shift correction to align the resonance peak of the spectral codes to that of the spectrum of the full open pattern. These wavelength shifts $\Delta \lambda$ are as small as $\sim$ \SI{3}{\nano\meter}, and are probably due to the fact that when measuring the spectra codes, the incident light is deviated from normal incidence. This deviation $\theta$ angle can be estimated by $\theta^2 = \Delta \lambda/\lambda^{peak}$, giving $\theta \approx 4^{\circ}$. 

\begin{figure}[ht!]
    \begin{center}
        \begin{tabular}{c}
        \includegraphics[width=16cm]{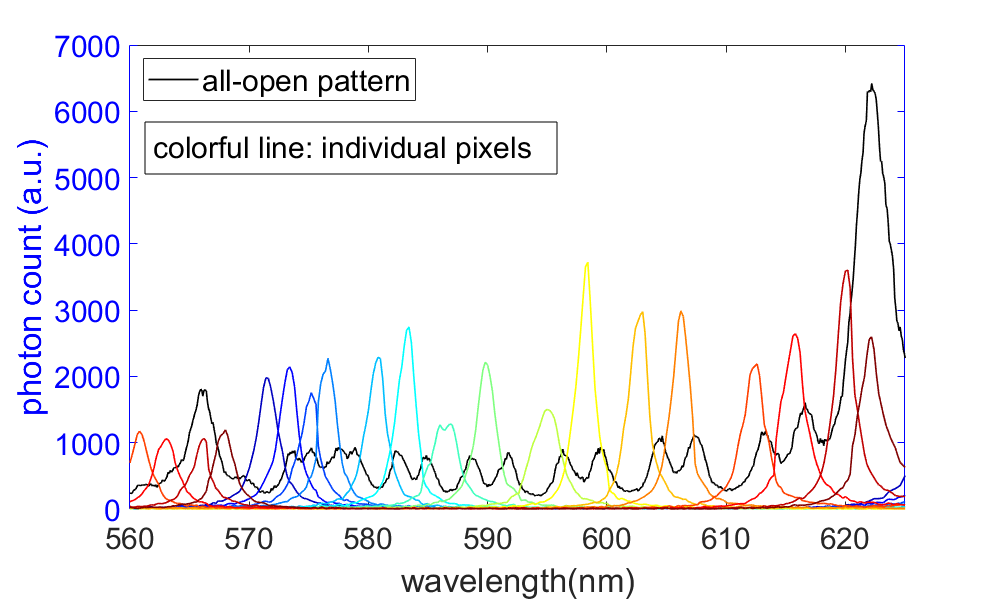}
        \end{tabular}
    \end{center}
    \caption{The measured transmission spectra of 16 different single pixels of the spectral encoder and an all-open pattern on a binary mask. The background is subtracted but the peak shift correction is not applied.} 
    \label{fgr:Fig_fullp_pixel}
\end{figure}



\newpage

\subsection*{V. Fabrication process flow}

\begin{figure}[ht!]
    \begin{center}
        \begin{tabular}{c}
        \includegraphics[width=16cm]{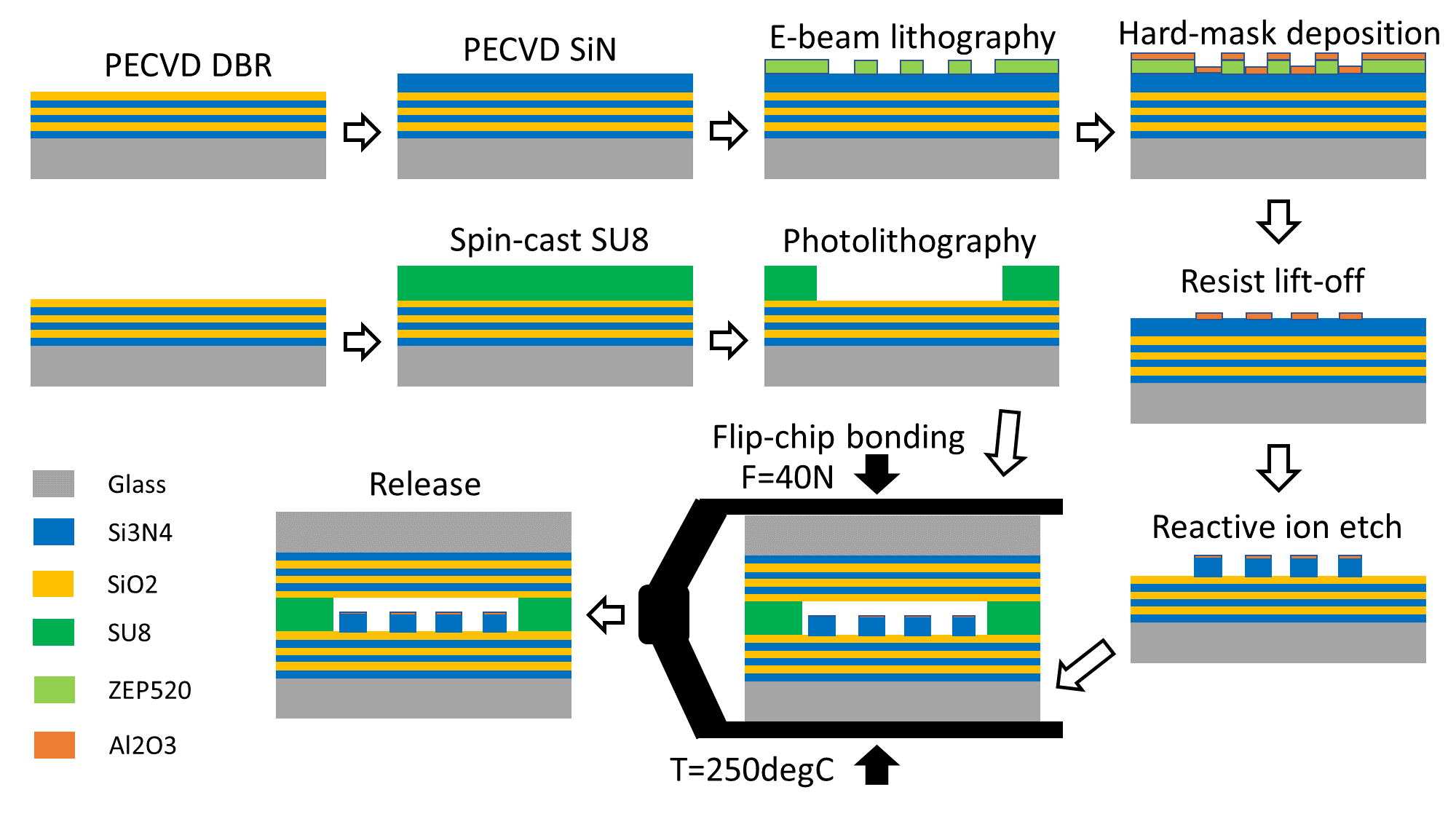}
        \end{tabular}
    \end{center}
    \caption{The sketch of the fabrication process flow of the SSE.}
    \label{fgr:Fig_processflow}
\end{figure}
\end{document}